\documentclass{article}
\usepackage{spconf,amsmath,amssymb,graphicx,booktabs}

\title{MULTI-DIMENSIONAL PROSODY JUDGMENT FOR \\
LIVE STREAMING SPEECH SYNTHESIS}

\name{Zifan Guan$^{1,2}$, \quad Longyu Lu$^{2}$, \quad Junan Zhang$^{1}$, \quad Zhizheng Wu$^{1}$, \quad Meiguang Jin$^{2}$\thanks{Corresponding author: Meiguang Jin.}, \quad Junfeng Ma$^{2}$}
\address{$^{1}$The Chinese University of Hong Kong, Shenzhen, China \\
$^{2}$TaoLive-AIGC Team \\
Taobao \& Tmall Group of Alibaba}

\begin{document}
\ninept
\maketitle

\begin{abstract}
Evaluating live streaming speech synthesis (TTS) requires assessing fine-grained,
highly expressive prosody---such as emotion, intonation, and energy---which
traditional MOS predictors fail to capture. While proprietary Large Language
Models (LLMs) like Gemini can evaluate these aspects, they are too costly for
massive inference and reinforcement learning feedback. To address this, we first
introduce Live-ProsodyJudge (LPJ), a cost-effective pairwise evaluator distilled
from Gemini into Qwen3-Omni.

However, we identify a critical flaw in standard multi-dimensional evaluation:
verdict coupling. The judge tends to lazily align all individual dimension scores
with its overall preference, collapsing a rich multi-dimensional rubric into a
single preference bit. To resolve this, we further propose
Decoupled-Live-ProsodyJudge (D-LPJ). D-LPJ eliminates the overall verdict target to
prevent blind following, masks uncertain pair-dimensions during Supervised
Fine-Tuning (SFT), and introduces a novel span-local GRPO strategy that applies
normalized advantages strictly to their corresponding rationale spans.

Evaluated on highly curated human-annotated test sets, 10-sample balanced-order LPJ
achieves higher point accuracy than a single Gemini call, while D-LPJ successfully
produces independent, decoupled dimension judgments. Furthermore, in a Best-of-8
TTS candidate selection tournament, the LPJ-selected utterance falls within the
human top-3 in 85.29\% of high-confidence cases, demonstrating its efficacy for
fine-grained TTS preference optimization.
\end{abstract}

\begin{keywords}
prosody evaluation, expressive speech evaluation, speech reward model,
reinforcement learning
\end{keywords}

\section{Introduction}
\label{sec:intro}
\begingroup
\setlength{\parskip}{0pt}

Live streaming TTS requires exceptional expressiveness: fluency, varied intonation,
appropriate emotion, high energy, and natural shifts between product explanation
and audience interaction. Traditional reference-free Mean Opinion Score (MOS)
predictors \cite{mosnet,utmos,nisqa} regress to naturalness or transmission quality,
inherently failing to capture these complex interaction-mode and expressive
phenomena.\par
Recent expressive and controllable TTS systems such as dots.tts
\cite{dotstts} make fine-grained evaluation increasingly important. Among
existing benchmarks, \mbox{EmergentTTS-Eval} \cite{emergentttseval} examines
complex prosodic, expressive, and linguistic challenges with model-as-a-judge, while
long-form speech benchmarking \cite{swanbench} covers diverse generation scenarios
and CEAEval \cite{ceaeval} tests expressive appropriateness in rich contexts.
SpeechJudge \cite{speechjudge} assesses speech naturalness, whereas GSRM
\cite{gsrm} and UniSRM \cite{unisrm} provide generative or reasoning-based
fine-grained rewards for speech RLHF. However, these methods do not target
live-streaming delivery and interaction.\par
Powerful audio-language models such as Gemini-3.1-pro-preview \cite{gemini} can make
difficult expressive judgments, but proprietary inference is prohibitively
expensive for iterative TTS evaluation and reinforcement learning optimization
(RLHF) over thousands of sampled utterances. We therefore distill Gemini's reasoning
into open-weight Qwen3-Omni \cite{qwen3omni}, establishing the baseline
Live-ProsodyJudge (LPJ) under a seven-dimension protocol: four always-scored core
dimensions (C1--C4)---fluency, intonation, emotion, and live streaming
expressiveness---and three transcript-triggered dimensions (L1--L3)---key-information
emphasis, emotional-state switching, and interaction-mode switching.\par
Despite achieving reasonable accuracy through swap-consistent distillation and
curriculum learning, we observe a severe systematic failure in the baseline LPJ:
verdict coupling. In a nominally multi-dimensional evaluation, the model often
collapses its judgments to follow the overall winner. For instance, if utterance A
is deemed better overall, the model lazily assigns wins to utterance A across
fluency, intonation, emotion, and expressiveness, severely limiting its utility for
targeted TTS optimization where distinct prosodic dimensions must yield distinct
verdicts.\par
To force dimension-independent evaluation, we introduce
Decoupled-Live-ProsodyJudge (D-LPJ). Because collapse is most severe in the four
dimensions scored for every pair, while the conditional L1--L3 dimensions provide
sparse supervision, D-LPJ focuses on C1--C4. It removes the overall-verdict target
and masks uncertain pair-dimensions during SFT rather than discarding the entire
pair. Finally, span-local GRPO \cite{grpo} applies each dimension's advantage only
to its own rationale span, preventing cross-dimension reward cancellation.\par
\noindent This work makes three main contributions:\par
\noindent\textbf{Protocol \& Data Formulation:} We define a multi-dimensional
prosody rubric for live streaming TTS and construct an evaluation suite of 1,043
internal human-annotated pairs (plus a 222-pair transfer set), strictly controlling
text transcripts to eliminate lexical confounding.\par
\noindent\textbf{Dimension-Decoupling Architecture:} We propose D-LPJ, which
mitigates verdict coupling by eliminating the overall-verdict target, employing
dimension-wise SFT masking, and introducing span-local GRPO for fine-grained credit
assignment.\par
\noindent\textbf{Empirical Validation:} Our results demonstrate that D-LPJ avoids
verdict collapse and yields varying multi-dimensional verdicts. Under 10-sample
balanced-order inference, LPJ exhibits negligible position bias and achieves higher
point agreement than a single Gemini call on four main test sets, while D-LPJ
achieves 86.10\% pooled agreement. In practical Best-of-8 candidate selection,
LPJ's top choice matches human top-3 preferences in 85.29\% of cases.\par
\endgroup
\section{Task and Data}
\label{sec:task}

For two clips $(a_A,a_B)$ rendering transcript $x$, the judge produces a rationale
$c_d$ and scores $(s_d^A,s_d^B)\in\{1,\ldots,10\}^2$ for dimension $d$. Its
per-dimension verdict is
\begin{equation}
 v_d=\operatorname{sgn}(s_d^A-s_d^B)\in\{A,\mathrm{Tie},B\}.
\end{equation}
Table \ref{tab:rubric} summarizes the seven dimensions. LPJ always evaluates
C1--C4, activates L1--L3 from transcript cues, and emits an overall verdict; D-LPJ
emits only C1--C4, with aggregation performed externally. All experiments use the
high-expressiveness profile.

\begin{table*}[t]
\centering
\ninept
\setlength{\tabcolsep}{4.0pt}
\caption{Seven-dimension live streaming prosody rubric. ``Text'' means that the
dimension is evaluated only when the transcript contains the corresponding cue.}
\label{tab:rubric}
\begin{tabular}{clcl}
\toprule
ID & Dimension & Gate & Principal attributes \\
\midrule
C1 & Fluency and Naturalness & Always & Rate stability, phrasing, pauses, coherence \\
C2 & Intonation Variation & Always & Pitch range, final contours, stress, sentence contrast \\
C3 & Emotional Expression & Always & Emotion--content match, intensity, warmth and appeal \\
C4 & Live Streaming Expressiveness & Always & Energy, rate drive, rhythm, real-host quality \\
L1 & Key Information Emphasis & Text & Prominence of price, discount, scarcity, and calls to action \\
L2 & Emotional State Switching & Text & Capture and smoothness of emotional turning points \\
L3 & Interaction Mode Switching & Text & Delivery change between explanation and audience address \\
\bottomrule
\end{tabular}
\end{table*}

The 1,043-pair main suite comprises Qwen3-TTS \cite{qwen3tts} base versus our SFT
variant (T1, 417; T1$\star$ is its 119-pair unanimous subset), two samples from our
internal Qwen3-TTS-GDPO variant (T2, 111 unanimous) testing within-model variation,
the same internal variant versus BERT-CosyVoice \cite{cosyvoice} (T3, 315), and
human recordings versus Qwen3-TTS (T4, 200). T5 adds 222 IndexTTS2
\cite{indextts2}-versus-FireRedTTS \cite{fireredtts} transfer pairs from a model
combination absent from coupled-judge training.

T1--T4 include 135 speakers across warm-recommendation, passionate-promotion,
and professional-explanation styles, with matched product categories. Transcripts
combine real-stream ASR and generated copy for product entry, explanation, and
ordering; durations are balanced across 5--10, 10--20, and 20--30 seconds. Each
pair shares one transcript, reducing lexical confounding across cross-model,
same-model, and human-versus-TTS comparisons.

Three trained contractors independently label each pair using the teacher's
rubric definitions and pairwise criteria, blinded to system and source. Across
T1--T4, 45\% of labels are unanimous 3--0 and 55\% have two agreeing votes and one
abstention; exclusions are limited to
corrupted audio, transcript mismatch, or protocol violations. Frozen evaluation
partitions do not overlap training or validation in audio, pair, normalized
transcript, speaker, or source stream. C1--C4 test sets contain 224, 172, 142, and
160 unanimous non-tie pairs. Audio is not redistributed because of licensing.

\section{Methodology}
\label{sec:method}

\subsection{Swap-consistent distillation and curriculum}

Gemini judges each training pair twice in both A/B orders; swapped votes are
mapped back to the clips. LPJ retains only 4--0 outcomes and balances the winner
between slots. Its curriculum moves from human-versus-TTS to cross- and same-model
comparisons. We attach rank-32 LoRA \cite{lora} to Qwen3-Omni attention projections
and merge it after SFT.

For section token set $T_d$ and dimension weight $\alpha_d$, the weighted SFT loss
is
\begin{equation}
 \mathcal L_{\mathrm{SFT}}=-\frac{\sum_d\alpha_d\sum_{t\in T_d}
 \log\pi_\theta(y_t\mid y_{<t},a_A,a_B,x)}
 {\sum_d\alpha_d|T_d|}.
\end{equation}
The coupled judge uses weights $(1.5,1.5,1.5,1.5,1.5,0.7,0.7)$, ordered as
(C1, C2, C3, C4, L1, L2, L3), with weights 1.0 and 2.0 on the summary and
conclusion, respectively.

\subsection{Dimension-wise supervision}

D-LPJ queries Gemini four times per core dimension, with each call returning only
that dimension's rationale and score pair. After restoring swapped votes, we retain
4--0, 3--1, or three agreeing non-tie votes plus one abstention. Other
pair-dimensions receive zero SFT loss while confident dimensions remain trainable.
Retention therefore occurs at the pair-dimension level instead of dropping an
entire pair when one label is uncertain. Retained sections form one C1--C4
rationale target without a summary or overall conclusion.

Both D-LPJ SFT stages use rank-64 QLoRA and learning rate $5\times10^{-5}$, for
8 and 14 epochs, respectively, with batch size 1 and eight-step gradient
accumulation.

For GRPO \cite{grpo}, $G=8$ completions are sampled for prompt $q$. For any
group-level signal $z_i$, we use
\begin{equation}
 \mathcal N_G(z_i)=\frac{z_i-\operatorname{mean}_{j}(z_j)}
 {\operatorname{std}_{j}(z_j)+\epsilon}.
\end{equation}
The coupled Live-ProsodyJudge receives one reward: $+1$ when its overall verdict
matches the target and $-1$ otherwise. Its normalized advantage is broadcast over
the full completion. For D-LPJ, external aggregation is
$V_w=\operatorname{sgn}[\sum_d w_d(s_d^A-s_d^B)]$, with zero treated as a tie. The
aggregate-reward baseline uses $w=(1,1,1,1)$ and broadcasts the resulting single
verdict reward over the completion, even though the model emits no conclusion.

Span-local GRPO instead defines, for every confident dimension,
\begin{equation}
 r_{i,d}=\begin{cases}+1,&v_{i,d}=\tilde v_d,\\-1,&\text{otherwise},\end{cases}
 \qquad d\in\{\mathrm{C1},\ldots,\mathrm{C4}\},
\end{equation}
and normalizes each dimension independently across the same rollout group. Its
token-level advantage is
\begin{equation}
 \hat A_{i,t}=\begin{cases}
 \mathcal N_G(r_{i,d}), & t\in T_{i,d}\text{ and }d\text{ is confident},\\
 0, & \text{otherwise}.
 \end{cases}
\end{equation}
Consequently, C1 reward cannot directly cancel C4 reward in the token objective,
although all dimensions share model parameters. If every rollout receives the same
reward for one dimension, normalization gives zero advantage and that dimension
contributes no policy gradient for the group. Both GRPO variants start from the
same two-stage D-LPJ SFT checkpoint and use a new rank-64 LoRA, learning rate
$10^{-5}$, reverse-KL coefficient 0.04, and three GRPO epochs. We add no format
reward.

\subsection{Decoupling design and evaluation}

D-LPJ reduces cross-dimension coupling in both targets and optimization.
Independent teacher queries generate dimension-specific rationales without an
overall verdict; pair-dimension masking retains reliable supervision even when
other dimensions of the same pair are uncertain. Removing the shared conclusion
keeps SFT focused on the four dimension outputs, while span-local GRPO assigns each
dimension's reward only to its own rationale. Together, these mechanisms encourage
distinct judgments and dimension-specific credit assignment.

Here, independent judgments refer to separate dimension-level decisions rather
than complete statistical independence: the four dimensions still share the same
audio, transcript, autoregressive context, and model parameters, and genuine
prosodic correlations may remain. We evaluate the resulting judgments in two
complementary settings. In the 140-pair multidimensional human set, three
annotators assess C1--C4 for every audio pair, with confidence determined separately
for each pair-dimension. The larger dimension-specific sets provide additional
high-confidence human labels for measuring accuracy on each individual axis.

\section{Experiments}
\label{sec:exp}

\subsection{Setup and coupled-judge results}

Models are trained with SWIFT \cite{swift} in bf16 on four 72GB GPUs. All reported
training comparisons use one fixed random seed. Gemini uses one judgment in the
presented order, reflecting its intended use and the cost of repeated proprietary
inference. SpeechJudge-GRM and our students use ten sampled judgments, five in each
A/B order. Swapped score pairs are restored to physical clips, each clip's scores
are averaged across the ten generations, and the verdict is the sign of the mean
score difference. A tie is counted as an error.

We report agreement with the final human A/B label. For paired system comparisons,
we draw 10,000 bootstrap resamples of test-pair indices and apply identical indices
to both systems; the 2.5th and 97.5th percentiles of the accuracy-difference
distribution define the 95\% interval. These intervals quantify test-pair sampling
uncertainty for the fixed checkpoints, not variation across training seeds.

\begin{table*}[t]
\centering
\ninept
\setlength{\tabcolsep}{4.5pt}
\caption{Overall human-label agreement (\%). ``1s'' denotes the first sampled
judgment and ``10s'' denotes 10-sample balanced-order aggregation; Gemini uses one
call. T1$\star$ is part of T1.}
\label{tab:main}
\begin{tabular}{lccccc}
\toprule
Model & T1 & T1$\star$ & T2 & T3 & T4 \\
\midrule
SpeechJudge-GRM (10s) & 55.88 & 72.27 & 48.65 & 53.97 & 42.00 \\
Gemini-3.1-pro-preview (1s) & 65.23 & 72.27 & 65.77 & 73.33 & 58.00 \\
Without stage-1 SFT (10s) & 53.00 & 72.27 & 60.36 & 69.84 & 45.50 \\
Single-stage SFT (10s) & 72.66 & 80.67 & 67.57 & 80.00 & 74.50 \\
LPJ v0, 4--0 + 3--1 (10s) & 71.22 & 78.99 & 65.77 & 80.63 & 73.00 \\
LPJ v1, 4--0 SFT (1s) & 64.99 & 74.79 & 63.06 & 73.97 & 66.50 \\
LPJ v1, 4--0 SFT (10s) & 70.74 & 80.67 & 70.27 & 82.22 & 80.00 \\
LPJ v1 + GRPO (1s) & 67.87 & 83.19 & 65.77 & 80.95 & 74.50 \\
LPJ v1 + GRPO (10s) & 71.22 & 82.35 & 70.27 & 82.86 & 83.50 \\
\bottomrule
\end{tabular}
\end{table*}

Table \ref{tab:main} shows that LPJ v1+GRPO has higher point accuracy than
single-call Gemini in every column. Relative to v1 SFT, its changes are
$+0.48,+1.68,0.00,+0.63,$ and $+3.50$ points on T1, T1$\star$, T2, T3, and T4.
Table \ref{tab:mainci} reports the paired intervals. Every interval includes zero,
so these are descriptive changes. Admitting lower-confidence 3--1 labels (v0) is
worse than 4--0-only v1 on four of five columns, indicating that confidence-filtered
supervision is more important than simply retaining more teacher-labeled data.
Removing the large-gap first stage is associated with a 34.5-point lower T4
estimate, while single-stage training is lower on T2--T4 but higher on T1.

\begin{table}[t]
\centering
\ninept
\setlength{\tabcolsep}{4pt}
\caption{Paired-bootstrap comparison of 10-sample LPJ v1+GRPO against v1 SFT.
Differences and 95\% intervals are in percentage points.}
\label{tab:mainci}
\begin{tabular}{lcc}
\toprule
Set & GRPO $-$ SFT & 95\% interval \\
\midrule
T1 & $+0.48$ & [$-2.64$, $+3.60$] \\
T1$\star$ & $+1.68$ & [$-2.52$, $+6.72$] \\
T2 & $0.00$ & [$-7.21$, $+7.21$] \\
T3 & $+0.63$ & [$-1.90$, $+3.18$] \\
T4 & $+3.50$ & [$-1.50$, $+8.50$] \\
\bottomrule
\end{tabular}
\end{table}

On T5, SpeechJudge-GRM obtains 185/222 (83.33\%), Gemini obtains 189/222 (85.14\%),
and LPJ v1+GRPO obtains 192/222 (86.49\%). The three-pair, 1.35-point gap over
Gemini has no paired confidence interval and is treated only as descriptive transfer
evidence.

\subsection{Position-bias diagnostic}

We measure position bias using all 4,170 judgments from the 417 T1 pairs under
balanced presentation. Each physical comparison occurs in both A/B assignments, so
pooling assignments separates presentation slot from clip identity. Let
$\Delta_{\mathrm{pos}}=\bar s_B-\bar s_A$ denote the difference between scores
assigned to the second- and first-presented slots. We also report the fraction of
raw judgments whose verdict favors presented B; the $t$ statistic is computed from
pair-level mean slot differences.

\begin{table}[t]
\centering
\ninept
\setlength{\tabcolsep}{2.2pt}
\caption{Position statistics on T1 under 10-sample balanced-order inference.}
\label{tab:posbias}
\begin{tabular}{lccccc}
\toprule
Model & $\bar s_A$ & $\bar s_B$ & $\Delta_{\mathrm{pos}}$ & $t$ & B win \\
\midrule
Live-ProsodyJudge & 7.842 & 7.854 & $+0.012$ & $+0.4$ & 50.7\% \\
SpeechJudge-GRM & 7.061 & 8.039 & $+0.978$ & $+11.7$ & 66.8\% \\
\bottomrule
\end{tabular}
\end{table}

Table \ref{tab:posbias} shows that SpeechJudge-GRM assigns the second-presented
clip nearly one extra point and favors B in two thirds of judgments. In contrast,
Live-ProsodyJudge is close to slot symmetry in both score gap and win rate. This is
consistent with swap-consistency filtering and winner-balanced training order, but
it is not a causal ablation of either component. Raw score levels are also not
compared across models; the diagnostic concerns within-model A/B asymmetry.

\subsection{Dimension-wise evaluation}

Across the evaluation sets, D-LPJ produces at least two different non-tie
core-dimension verdicts on 24.3--60.6\% of pairs, showing that it can avoid
unanimous verdict vectors. We additionally use a 140-pair multidimensional human
test set. For each audio pair, three trained annotators independently label C1--C4;
confidence is determined separately for every pair-dimension, so not every pair has
confident labels on all four axes. On this set, the coupled LPJ produces 0/140
non-unanimous core vectors and 48.0\% agreement on the confident human dimension
labels. D-LPJ SFT produces 90/140 non-unanimous vectors and 67.9\% agreement;
span-local GRPO changes these to 100/140 (71.4\%) and 71.4\%, respectively. This
same-set comparison evaluates both resistance to verdict collapse and agreement
with the retained human judgments.

Table \ref{tab:perdim} therefore evaluates each dimension against a separate,
unanimous human test set. Ten-sample SFT exceeds one-call Gemini in every dimension.
Span-local GRPO changes pooled agreement from 73.64\% to 77.94\% with one sample and
from 84.10\% to 86.10\% with ten samples. At ten samples it is 1.00 point above
aggregate-reward GRPO. Table \ref{tab:dimci} shows that all per-dimension paired
intervals for span-local GRPO versus SFT and aggregate-reward GRPO include zero; we
therefore report point estimates without claiming statistical significance.

\begin{table}[t]
\centering
\ninept
\setlength{\tabcolsep}{1.4pt}
\caption{D-LPJ agreement with unanimous human dimension labels (\%). ``Agg.''
and ``Span'' denote aggregate-reward and span-local GRPO.}
\label{tab:perdim}
\begin{tabular}{lccccc}
\toprule
Model & C1 & C2 & C3 & C4 & All \\
\midrule
Gemini (1s) & 74.55 & 77.33 & 80.28 & 84.38 & 78.65 \\
SFT (1s) & 68.75 & 65.70 & 76.06 & 86.88 & 73.64 \\
SFT (10s) & 79.02 & 80.81 & 85.92 & 93.13 & 84.10 \\
Agg. GRPO (1s) & 70.54 & 71.51 & 79.58 & 90.62 & 77.22 \\
Agg. GRPO (10s) & 79.91 & 83.14 & 87.32 & 92.50 & 85.10 \\
Span GRPO (1s) & 71.88 & 76.74 & 76.76 & 88.75 & 77.94 \\
Span GRPO (10s) & 83.04 & 83.14 & 85.21 & 94.38 & 86.10 \\
\bottomrule
\end{tabular}
\end{table}

\begin{table}[t]
\centering
\ninept
\setlength{\tabcolsep}{2.5pt}
\caption{Ten-sample span-local GRPO differences and paired-bootstrap 95\%
intervals, in percentage points.}
\label{tab:dimci}
\begin{tabular}{lcc}
\toprule
Dim. & Versus SFT & Versus Agg. GRPO \\
\midrule
C1 & $+4.02$ [$-0.89$, $+9.38$] & $+3.12$ [$-1.79$, $+7.59$] \\
C2 & $+2.33$ [$-2.33$, $+6.98$] & $0.00$ [$-5.23$, $+5.23$] \\
C3 & $-0.70$ [$-4.23$, $+2.82$] & $-2.11$ [$-6.34$, $+2.11$] \\
C4 & $+1.25$ [$-1.88$, $+4.38$] & $+1.88$ [$-1.88$, $+5.62$] \\
\bottomrule
\end{tabular}
\end{table}

\subsection{Additional diagnostics}

Repeated, balanced-order inference contributes unevenly. Single-sample LPJ
v1+GRPO obtains 67.87, 83.19, 65.77, 80.95, and 74.50\% on T1, T1$\star$, T2, T3,
and T4; ten-sample aggregation changes these estimates by $+3.35$, $-0.84$,
$+4.50$, $+1.91$, and $+9.00$ points. It combines repeated generation with order
balancing, whose effects are not isolated here.

D-LPJ also permits external scalar decisions without restoring an overall text
target. Validation-selected weights $w=(3,0.5,0.5,3)$ give 73.6\% on T1 and 85.5\%
on T4, versus 69.3\% and 78.5\% with uniform weights; T2 remains 71.2\%. This is
not a causal comparison with the coupled judge because their objectives differ.

We also test CEAEval's adaptive audio-attention bias \cite{ceaeval}. A representative
instance contains 426 audio, 634 rationale, and 2,618 rubric-prompt tokens. It
trains stably but does not improve held-out accuracy, so it is excluded from the
final model. One possible explanation is that the generated rationale is not much
longer than the audio sequence, limiting the benefit of explicitly up-weighting
audio attention; substantially longer generations may behave differently.

\subsection{Best-of-8 and limitations}

For each transcript, the TTS model generates eight candidates and
Live-ProsodyJudge evaluates all $\binom{8}{2}=28$ distinct pairs. Without requiring
absolute-score calibration, we select the candidate with the largest number of
pairwise wins,
\begin{equation}
 \hat a=\arg\max_{a_i}\sum_{j\ne i}\mathbf 1[V(a_i,a_j)=a_i].
\end{equation}
Two trained contractor annotators independently rank the top three candidates for
400 sets. We retain the 136 sets for which both produce the same ordered top-three
ranking; this inclusion decision is made without model predictions. On this
high-confidence subset, the selector obtains Hit@1/2/3 of 98/136 (72.06\%),
106/136 (77.94\%), and 116/136 (85.29\%), compared with random-selection baselines
of 12.5\%, 25.0\%, and 37.5\%. Hit@$k$ means that the tournament winner lies within
the agreed human top-$k$. These rates characterize the retained subset and do not
estimate performance on all 400 collected sets.

Three limitations bound the evidence. First, training results use one seed;
bootstrap intervals quantify test-pair sampling uncertainty, not variation across
training runs. Second, ten-sample comparisons against Gemini use more inference
and balanced order. Third, although the 140-pair multidimensional set provides
complete human C1--C4 annotations, most pairs in the larger dimension-specific test
sets do not include complete annotations across all four dimensions. D-LPJ is
currently limited to C1--C4, and using its vector rewards to optimize a TTS model
remains future work.

\section{Conclusion}
\label{sec:conclusion}

We presented LPJ and D-LPJ for multi-dimensional live streaming prosody evaluation.
Balanced-order LPJ achieves strong point agreement, reduces measured slot bias
relative to SpeechJudge-GRM, and supports Best-of-8 selection. D-LPJ removes the
shared overall target and localizes SFT and GRPO supervision by dimension, producing
varied core-dimension judgments with 86.10\% pooled agreement. Together, the two
judges provide scalable candidate ranking and dimension-specific diagnosis. Future
work will extend decoupling to sparse conditional dimensions and use dimension-wise
rewards for TTS post-training.

\section{Funding Acknowledgment}
This work was supported by the TaoLive-AIGC Team, Taobao \& Tmall Group of Alibaba.

\section{Compliance with Ethical Standards}
Human labels were produced by contracted annotation staff who received
task-specific training and were blinded to system identity. We report only
aggregate results. Evaluation audio is not publicly redistributed because source
recordings are subject to copyright and licensing restrictions.

\end{document}